\documentclass[a4paper]{jpconf}
\usepackage{graphicx}
\usepackage{amssymb}
\begin{document}
\title{A frequency-domain implementation of the particle-without-particle approach to EMRIs}

\author{
Marius Oltean$^{1,2,3,4,5}$, 
Carlos F Sopuerta$^{1}$ 
and Alessandro D A M Spallicci$^{3,4,5}$}

\address{$^{1}$ {\mbox Institut de Ci\`encies de l'Espai (CSIC-IEEC), Campus
Universitat Aut\`{o}noma de Barcelona}\\
{\mbox  Carrer de Can Magrans s/n, 08193 Cerdanyola del Vall\`es, Spain}}

\address{$^{2}$ {\mbox Departament de F\'isica, Facultat de Ci\`{e}ncies,
Universitat Aut\`{o}noma de Barcelona}\\
{\mbox Edifici C, 08193 Cerdanyola del Vall\`{e}s, Spain}}

\address{$^{3}$ {\mbox Observatoire des Sciences de l'Univers en r\'{e}gion Centre
(OSUC), Universit\'{e} d'Orl\'{e}ans}\\
{\mbox 1A rue de la F\'{e}rollerie, 45071 Orl\'{e}ans, France}}

\address{$^{4}$ {\mbox P\^{o}le de Physique, Collegium Sciences et Techniques
(CoST), Universit\'{e} d'Orl\'{e}ans} \\
{\mbox Rue de Chartres, 45100  Orl\'{e}ans, France}}

\address{$^{5}$ {\mbox Laboratoire de Physique et Chimie de l'Environnement et de
l'Espace (LPC2E)}\\
{\mbox Centre National de la Recherche Scientifique (CNRS)}\\
{\mbox 3A Avenue de la Recherche Scientifique, 45071 Orl\'{e}ans, France}}

\ead{oltean@ice.cat}

\begin{abstract}
The gravitational waves emitted by binary systems with extreme mass ratios carry unique astrophysical information expected to be probed by the next generation of gravitational wave detectors such as LISA. The detection of these binaries rely on an accurate modeling of the gravitational self-force that drives their orbital evolution. Although the theoretical formalism to compute the self-force has been largely established, the mathematical tools needed to implement it are still under development, and the self-force computation remains an open problem. We present here a frequency-domain implementation of the particle-without-particle (PwP) technique previously developed for the computation of the scalar self-force -- a helpful testbed for the gravitational self-force.
\end{abstract}

\section{Introduction and Motivation}

Extreme-Mass-Ratio Inspirals (EMRIs) are one of the main sources of gravitational waves (GWs) for space-based detectors like the LISA mission. EMRIs  are binary system which consist of a stellar compact object (SCO; with a mass range $m_{\ast}\sim 1-50M_{\odot}$) orbiting a massive black hole (MBH; with a mass range $M_{\bullet}\sim 10^{5-7}M_{\odot}$).  In the regime where the dynamics is driven by GW emission, the SCO inspirals into the MBH sweeping through the LISA frequency band, and mapping the MBH spacetime onto the structure of the GWs in great detail.  However, due to the complexity of EMRI GW signals, we need precise theoretical waveform templates to extract the physical parameters of the system from the detector data stream.

The challenge in modeling EMRIs is to compute the perturbations generated by the SCO in the (background) gravitational field of the MBH, and how these perturbations affect the motion of the SCO itself. The most extended approach \cite{Poisson} consists in modelling the SCO by using a point-like description, and then, to describe the backreaction effects on the dynamics as the action of a local {\em self-force} that is responsible for the deviations from geodesic motion.

While tremendous progress has been made on this problem in recent years, significant obstacles remain -- stemming, in large part, from the singular nature of the gauge where the metric perturbations are usually computed, which renders the self-force calculation computationally challenging. To the effect of developing and testing an effective method for circumventing these issues in the full gravitational case, we consider here the simpler problem of the self-force due to a scalar field in the frequency domain, via a previously developed technique called the particle-without-particle (PwP) approach \cite{Canizares2009,Canizares2011}; for other similar approaches using jump conditions, see
\cite{aoudiaspallicci2011}.

\section{Setup}
In our simplified EMRI model, the SCO is a charged scalar particle, with charge $q$ associated to a scalar field $\Phi$, orbiting a non-rotating MBH with a fixed geometry, the Schwarzschild-Droste metric. Because of spherical symmetry in this background, the field harmonic modes $\Phi^{\ell m}(t,r)$ decouple, and the scalar field equation reduces to wave-like PDEs for $\psi^{\ell m}=r\Phi^{\ell m}$,
\begin{equation}
\left(\square -V^{}_{\ell}(r) 
\right)\psi^{\ell m}= S^{\ell m}\delta (r-r^{}_{p}(t))\,, 
\label{master}
\end{equation}
where $\square = -\partial^2_{t} + \partial^2_{r_{*}}$, $r_{*}=r+2M_{\bullet}\ln(r/2M_{\bullet}-1)$, $V_{\ell}(r)$ is the so-called Regge-Wheeler potential, and $S^{\ell m}$ is a source term proportional to $q$ and dependent on the particle location $r^{}_{p}(t)$. Once the field is solved for, its singular part must be subtracted (via ``mode-sum regularisation''), leaving us with its ``regular'' part from which the value of the self-force can be calculated \cite{Poisson}.

The full solution has to be found numerically, but the presence of singularities makes the task difficult. To circumvent this, the idea of the PwP method is to split the computational domain into two disjoint regions whereby any non-singular quantity ${\cal Q}(t,r)$ is decomposed as ${\cal Q} = {\cal Q}_{-}\Theta^{-}_{p} + {\cal Q}_{+}\Theta^{+}_{p}$, where $\Theta^{\pm}_{p}= \Theta(\pm(r - r_p))$ is the Heaviside step function. Quantities that are not continuous will have jumps across the SCO trajectory: $[{\cal Q}]_p = \lim_{r\rightarrow r^{}_{p}(t)} {\cal Q}_{+}(t,r) - {\cal Q}_{-}(t,r)$. In this setup then, the problem (\ref{master}) with a singular source is effectively replaced with homogeneous equations to the left and right of the SCO, subject to certain jump conditions across it.

\section{Frequency-domain numerical implementation and results}

Previously, the PwP method has been used to solve this problem (as a PDE) directly in the time domain.
But this approach generically leads to slow computational times, and may not be adaptable to more realistic spacetimes such as Kerr. To ameliorate these issues, we can tackle the problem in the frequency domain: we expand the fields in discrete Fourier series, transforming the scalar field PDEs into Schr\"{o}dinger equation-like ODEs -- supplemented by appropriate (non-reflecting) boundary conditions as well as the jump conditions.

We use a pseudospectral collocation method to find numerical solutions for the Fourier modes, which are then matched by imposing the jump conditions. Thus far, using this approach, the known value of the self-force has been recovered for circular orbits (in agreement with the PwP in the time domain and other methods in the literature); we are working on extending this to generic (eccentric) orbits, and we aim to also adapt this method to the Kerr spacetime.

\paragraph*{\bf Acknowledgements}
Due by MO to the Natural Sciences and Engineering Research Council of Canada, 
by MO and ADAMS to LISA France-CNES, and by CFS to the Ministry of Economy and
Competitivity of Spain, MINECO, contracts ESP2013-47637-P and ESP2015-67234-P.

\end{document}